

\def\Titlehh#1#2{\nopagenumbers\abstractfont\hsize=\hstitle\rightline{#1}%
\vskip .2in\centerline{\titlefont #2}\abstractfont\vskip .2in\pageno=0}
\def\newsecnpb#1{\global\advance\secno by1\message{(\the\secno. #1)}
\global\subsecno=0\eqnres@t\centerline{\bf\the\secno. #1}
\writetoca{{\secsym} {#1}}\par\nobreak\medskip\nobreak}
\def\eqnres@t{\xdef\secsym{\the\secno.}\global\meqno=1\bigbreak\bigskip}
\def\sequentialequations{\def\eqnres@t{\bigbreak}}\xdef\secsym{}
\global\newcount\subsecno \global\subsecno=0
\def\subsecnpb#1{\global\advance\subsecno by1
\message{(\secsym\the\subsecno. #1)}
\ifnum\lastpenalty>9000\else\bigbreak\fi
\noindent{\secsym\the\subsecno. #1}\writetoca{\string\quad
{\secsym\the\subsecno.} {#1}}\par\nobreak\medskip\nobreak}

\def\CTPa{\it Center for Theoretical Physics, Department of Physics,
      Texas A\&M University}
\def\CTPb{\it College Station, TX 77843-4242, USA}
\def\HARCa{\it Astroparticle Physics Group,
Houston Advanced Research Center (HARC)}
\def\HARCb{\it The Woodlands, TX 77381, USA}

\def\CERN{\it CERN Theory Division, 1211 Geneva 23, Switzerland}
\def\ie{\hbox{\it i.e.}}     
\def\eg{\hbox{\it e.g.}}     

\def\nextline{\unskip\nobreak\hfill\break}
\def\coeff#1#2{{\textstyle{#1\over #2}}}

\catcode`\@=11 

\def\lsim{\mathrel{\mathpalette\@versim<}}
\def\gsim{\mathrel{\mathpalette\@versim>}}
\def\@versim#1#2{\vcenter{\offinterlineskip
    \ialign{$\m@th#1\hfil##\hfil$\crcr#2\crcr\sim\crcr } }}
\def\boxit#1{\vbox{\hrule\hbox{\vrule\kern3pt
      \vbox{\kern3pt#1\kern3pt}\kern3pt\vrule}\hrule}}

\def\r#1{$\bf#1$}
\def\rb#1{$\bf\overline{#1}$}

\def\t1{{\tilde 1}}
\def\ov{\overline}

\def\JL{J. L. Lopez}
\def\DVN{D. V. Nanopoulos}

\def\GeV{\,{\rm GeV}}
\def\TeV{\,{\rm TeV}}

\def\wt{\widetilde}

\def\NPB#1#2#3{Nucl. Phys. B {\bf#1} (19#2) #3}
\def\PLB#1#2#3{Phys. Lett. B {\bf#1} (19#2) #3}

\def\PRD#1#2#3{Phys. Rev. D {\bf#1} (19#2) #3}
\def\PRL#1#2#3{Phys. Rev. Lett. {\bf#1} (19#2) #3}
\def\PRT#1#2#3{Phys. Rep. {\bf#1} (19#2) #3}

\def\TAMU#1{Texas A \& M University preprint CTP-TAMU-#1}

\nref\Barr{S. Barr, \PLB{112}{82}{219}, \PRD{40}{89}{2457}; J. Derendinger,
J. Kim, and \DVN, \PLB{139}{84}{170}.}
\nref\revitalized{I. Antoniadis, J. Ellis, J. Hagelin, and \DVN,
\PLB{194}{87}{231}.}
\nref\revamp{I. Antoniadis, J. Ellis, J. Hagelin, and \DVN,
\PLB{231}{89}{65}.}
\nref\faspects{J. Ellis, J. Hagelin, S. Kelley, and \DVN, \NPB{311}{88/89}{1}.}
\nref\Lacaze{I. Antoniadis, J. Ellis, R. Lacaze, and \DVN, \PLB{268}{91}{188};
S. Kalara, \JL, and \DVN, \PLB{269}{91}{84}.}
\nref\msu{J. Ellis, S. Kelley, and \DVN, \PLB{249}{90}{441};
F. Anselmo, L. Cifarelli, and A. Zichichi, CERN-PPE/92-145 and
CERN/LAA/MSL/92-011 (July 1992).}
\nref\price{I. Antoniadis, J. Ellis, S. Kelley, and \DVN, \PLB{272}{91}{31};
D. Bailin and A. Love, \PLB{280}{92}{26}.}
\nref\sism{S. Kelley, \JL, and \DVN, \PLB{278}{92}{140}; G. Leontaris,
\PLB{281}{92}{54}.}
\nref\jhreview{For a recent review see \eg, \JL\ and \DVN, in Proceedings of
the 15th Johns Hopkins Workshop on Current Problems in Particle Theory, August
1991, p. 277; ed. by G. Domokos and S. Kovesi-Domokos.}
\nref\EKN{J. Ellis, S. Kelley, and \DVN, \PLB{260}{91}{131}.}
\nref\decisive{J. L. Lopez and \DVN, \PLB{251}{90}{73}, \PLB{256}{91}{150},
and \PLB{268}{91}{359}.}
\nref\LN{For a review see A. B. Lahanas and D. V. Nanopoulos,
\PRT{145}{87}{1}.}
\nref\Witten{E. Witten, \PLB{155}{85}{151}.}
\nref\nsI{J. Ellis, C. Kounnas, and \DVN, \NPB{241}{84}{406}.}
\nref\nsII{J. Ellis, C. Kounnas, and \DVN, \NPB{247}{84}{373}.}
\nref\sample{J. Ellis, A. Lahanas, \DVN, and K. Tamvakis, \PLB{134}{84}{429}.}
\nref\noscale{S. Kelley, \JL, \DVN, H. Pois, and K. Yuan, \PLB{273}{91}{473}.}
\nref\KLN{S. Kalara, J. Lopez, and \DVN, \PLB{245}{90}{421},
\NPB{353}{91}{650}.}
\nref\EVA{S. Kelley, \JL, and \DVN, \PLB{261}{91}{424}.}
\nref\chorus{J. Ellis, \JL, and \DVN, \PLB{292}{92}{189}.}
\nref\search{\JL, \DVN, K. Yuan, \TAMU{11/92}.}
\nref\dfive{M. Matsumoto, J. Arafune, H. Tanaka, and K. Shiraishi,
University of Tokyo preprint ICRR-267-92-5 (April 1992);
R. Arnowitt and P. Nath, \PRL{69}{92}{725};
J. Hisano, H. Murayama, and T. Yanagida, \PRL{69}{92}{1014} and
Tohoku University preprint TU--400 (July 1992).}
\nref\LNZ{\JL, \DVN, and A. Zichichi,
\PLB{291}{92}{255}; \JL, \DVN, and H. Pois, \TAMU{61/92} and CERN-TH.6628/92.}
\nref\Bethke{S. Bethke, Talk given at the XXVI International Conference
on High Energy Physics, Dallas, August 1992.}
\nref\aspects{S. Kelley, \JL, \DVN, H. Pois, and K. Yuan, \TAMU{16/92} and
CERN-TH.6498/92.}
\nref\ANc{P. Nath and R. Arnowitt, \PLB{289}{92}{368}.}
\nref\DNh{M. Drees and M. M. Nojiri, \PRD{45}{92}{2482}.}
\nref\higgs{S. Kelley, \JL, \DVN, H. Pois, and K. Yuan, \PLB{285}{92}{61}.}
\nref\LNY{\JL, \DVN, and K. Yuan, \NPB{370}{92}{445} and \PLB{267}{91}{219};
S. Kelley, \JL, \DVN, H. Pois, and K. Yuan, \TAMU{56/92} and CERN-TH.6584/92.}
\nref\KT{See \eg, E. Kolb and M. Turner, {\it The Early Universe}
(Addison-Wesley, 1990).}
\nref\many{R. Schaefer and Q. Shafi, Bartol preprint BA-92-28 (1992);
G. Efstathiou, J. R. Bond, and S. D. M. White, Oxford University preprint
OUAST/92/11; A. N. Taylor and M. Rowan-Robinson, Queen Mary College preprint,
June 1992.}
\nref\conden{S. Kalara, \JL, and \DVN, \PLB{275}{92}{304}.}
\nref\ART{I. Antoniadis, J. Rizos, and K. Tamvakis, \PLB{278}{92}{257}.}

\nfig\I{The running of the gauge couplings in the flipped $SU(5)$ model
for $\alpha_3(M_Z)=0.118$ (solid lines). The gap particle masses have been
derived using the gauge coupling RGEs to achieve unification at
$M_U=10^{18}\GeV$. The case with no gap
particles (dotted lines) is also shown; here $M_U\approx10^{16}\GeV$.}
\nfig\II{Scatter plots of the sbottom ($\tilde b_{1,2}$) and the stop
($\tilde t_{1,2}$) squark mass eigenstates versus $m_{\tilde g}$. (The $\tilde
b_{1,2}$ cannot be resolved in this manner.) The average squark mass (first two
generations) is $m_{\tilde q}\approx0.97m_{\tilde g}$. From this figure and on
all results are shown for both signs of $\mu$.}
\nfig\III{Scatter plots for the stau ($\tilde\tau_{1,2}$) and selectron
(or smuon) ($\tilde e_{L,R}$) masses. Note the spread in the $\tilde\tau_{1,2}$
masses for fixed $m_{\tilde g}$, due to the off-diagonal entries in the stau
mass matrix. The $\tilde\nu$ mass (not shown) starts off slightly below the
$\tilde e_R$ mass and then quickly joins the $\tilde e_L$ line.}
\nfig\IV{Scatter plots of the Higgs mixing term ($\mu$) versus $m_{\tilde g}$
and the neutralino cosmic relic abundance ($\Omega_\chi h^2_0$) versus
$m_\chi$. Note the proportionality
$|\mu|\propto m_{\tilde g}$ whose slope increases with $m_t$ ($m_t=100,130,160
\GeV$ shown). Also, for $m_\chi\approx{1\over2}M_Z$, the $Z$-pole
annihilation is quite noticeable.}
\nfig\V{Scatter plots of the second-to-lightest neutralino ($\chi^0_2$) mass
versus the lightest chargino ($\chi^\pm_1$) mass and second-to-lightest
($\chi^0_2$) to lightest ($\chi^0_1$) neutralino masses. Note the accuracy of
the $m_{\chi^0_2}=m_{\chi^\pm_1}$ relation.}
\nfig\VI{Scatter plots of the one-loop corrected lightest Higgs boson mass
$m_h$ versus the gluino mass for $m_t=100,130,160\GeV$ (top row), and the
pseudoscalar Higgs mass $m_A$ versus $m_h$ (bottom row). The three noticeable
bands (from
bottom-to-top in the top row and from left-to-right in the bottom row)
correspond to $m_t=100,130,160\GeV$. Note that in this model $m_A>m_h$
always. The heavy Higgs boson masses $m_H$ and $m_{H^+}$ are approached quickly
from below by $m_A$.}
\nfig\VII{Central values for the sparticle and one-loop corrected Higgs
boson masses for $m_{\tilde g}=300\GeV$, $m_t=130\GeV$, $\alpha_3(M_Z)=0.118$,
and (a) $\mu>0$ (b) $\mu<0$. The masses generally scale with $m_{\tilde g}$.
The percentage deviations from the shown values due to the variations in
$\tan\beta$ are given in Table III.}
\nfig\VIII{Scatter plots of $\tan\beta$ versus $m_{\tilde g}$ for the strict
no-scale case (where $B(M_U)=0$) for the indicated values of $m_t$. Note that
the sign of $\mu$ is {\it determined} by $m_t$ and that $\tan\beta$ can be
double-valued for $\mu>0$. Also shown are the values of the neutralino relic
abundance ($\Omega_\chi h^2_0$ versus $m_\chi$) for the same values of the
parameters.}
\nfig\IX{Same as Fig. 6, but for the strict no-scale case. Note the
weak dependence of $m_h$ on the gluino mass $m_{\tilde g}$. Also, if $\mu>0$
for $m_t\lsim135\GeV$, $m_h\lsim105\GeV$; whereas if $\mu<0$, for
$m_t\gsim140\GeV$, $m_h\gsim100\GeV$.}

\Titlehh{\vbox{\baselineskip12pt\hbox{CERN-TH.6667/92}\hbox{CTP--TAMU--68/92}
\hbox{ACT--20/92}}}
{\vbox{\centerline{The Simplest, String-Derivable, Supergravity Model}
\centerline{and its Experimental Predictions}}}
\centerline{JORGE~L.~LOPEZ$^{(a)(b)}$, D.~V.~NANOPOULOS$^{(a)(b)(c)}$,
and A. ZICHICHI$^{(d)}$}
\smallskip
\centerline{$^{(a)}$\CTPa}
\centerline{\CTPb}
\centerline{$^{(b)}$\HARCa}
\centerline{\HARCb}
\centerline{$^{(c)}$\CERN}
\centerline{$^{(d)}${\it CERN, Geneva, Switzerland}}
\vskip .1in
\centerline{ABSTRACT}
We present the simplest, string-derivable, supergravity model and discuss its
experimental consequences. This model is a new string-inspired flipped $SU(5)$
which unifies at the string scale $M_U=10^{18}\GeV$ due to the introduction of
an additional pair of \r{10},\rb{10} flipped $SU(5)$ representations which
contain new intermediate scale `gap' particles. We study various model-building
issues which should be addressed in string-derived incarnations of this model.
We focus our study on the no-scale supergravity mechanism and explore
thoroughly the three-dimensional parameter space of the model ($m_{\tilde
g},m_t,\tan\beta$), thus obtaining several simple relationships among the
particle masses, such as: $m_{\tilde q}\approx m_{\tilde g}$, $m_{\tilde
e_L}\approx m_{\tilde\nu}\approx0.30 m_{\tilde g}$, $m_{\tilde
e_R}\approx0.18m_{\tilde g}$, and
$m_{\chi^0_2}\approx2m_{\chi^0_1}\approx m_{\chi^\pm_1}$. In a strict
interpretation of the no-scale supergravity scenario we solve for $\tan\beta$
as a function of $m_t$ and $m_{\tilde g}$, and show that $m_t$ determines not
only the sign of the Higgs mixing parameter $\mu$ but also whether the lightest
Higgs boson mass is above or below $100\GeV$. We also
find that throughout the parameter space the neutralino relic abundance is
within observational bounds ($\Omega_\chi h^2_0\lsim0.25$) and may account
for a significant portion of the dark matter in the Universe.
\bigskip
\bigskip
{\vbox{\baselineskip12pt\hbox{CERN-TH.6667/92}\hbox{CTP--TAMU--68/92}
\hbox{ACT--20/92}}}
\Date{September, 1992}

\newsecnpb{Introduction}
The purpose of this paper is to find out the simplest supergravity model
compatible with all boundary conditions imposed by present experimental and
theoretical knowledge. The first property of this model is the number of
parameters needed, which we restrict to a minimum. In this search we follow
string-inspired choices. The most significant is the ``no-scale" supergravity
condition which -- in addition to being the only known mechanism to guarantee
the existence of light supersymmetric particles -- has the very interesting
property of being the
infrared limit of superstring theory. The other choices, aimed at the minimum
number of free parameters, are at present inspired by string phenomenology
and are good candidates to being rigorously derivable from string theory.
Our main goal is to produce a model whose basic conceptual choices are
attractive, in terms of what we think (and hope) will be the final theory of
all particles and interactions. One point needs to be emphasized. In order to
put string theory under experimental test, the first step is to construct
models with a number of parameters, which is as minimal as possible. Our aim
is to propose experimental tests that are steps towards the inclusion or
exclusion of our choices needed to build the model.

Besides the very economic GUT symmetry breaking mechanism in flipped $SU(5)$
\refs{\Barr,\revitalized} -- which allows it to be in principle derivable from
superstring theory \revamp\ -- perhaps one of the more interesting motivations
for considering such a unified gauge group is the natural avoidance of
potentially
dangerous dimension-five proton decay operators \faspects. In this paper we
construct a supergravity model based on this gauge group, which has the
additional property of unifying at a scale $M_U={\cal O}(10^{18})\GeV$, as
expected to occur in string-derived versions of this model \Lacaze.
As such, this model should constitute a blueprint for string model builders.
This string unification scale should be contrasted with the naive unification
scale, $M_U={\cal O}(10^{16}\GeV)$, obtained by running the Standard Model
particles and their superpartners to very high energies. This apparent
discrepancy of two orders of magnitude \msu\ creates a {\it gap} which needs to
be bridged somehow in string models. It has been shown \price\ that the
simplest solution to this problem is the introduction in the spectrum of heavy
vector-like particles with Standard Model quantum numbers. The
minimal such choice \sism, a quark doublet pair $Q,\bar Q$ and a $1/3$--charge
quark singlet pair $D,\bar D$, fit snugly inside a \r{10},\rb{10} pair of
flipped $SU(5)$ representations, beyond the usual
$3\cdot({\bf10}+\ov{\bf5}+{\bf1})$ of matter and \r{10},\rb{10} of Higgs.

In this model, gauge symmetry breaking occurs due to vacuum expectation values
(vevs) of the neutral components of the \r{10},\rb{10} Higgs representations,
which develop along flat directions of the scalar potential. There are two
known ways in which these vevs (and thus the symmetry breaking scale) could
be determined:\nextline
\indent (i) In the conventional way, radiative corrections to the scalar
potential in the presence of soft supersymmetry breaking generate a global
minimum of the potential for values of the vevs slightly below the scale
where supersymmetry breaking effects are first felt in the observable sector
\faspects. If the latter scale is the Planck scale (in a suitable
normalization)
then $M_U\sim M_{Pl}/\sqrt{8\pi}\sim 10^{18}\GeV$.\nextline
\indent (ii) In string-derived models a pseudo $U_A(1)$ anomaly arises as a
consequence of truncating the theory to just the massless degrees of freedom,
and adds a contribution to its $D$-term, $D_A=\sum q^A_i|\vev{\phi_i}|^2
+\epsilon$, with $\epsilon=g^2{\rm Tr\,}U_A(1)/192\pi^2\sim(10^{18}\GeV)^2$
\jhreview. To avoid a huge breaking of supersymmetry we need to demand $D_A=0$
and therefore the fields charged under $U_A(1)$ need to get suitable vevs.
Among these one generally finds the symmetry breaking Higgs fields, and
thus $M_U\sim10^{18}\GeV$ follows.\nextline
\indent In general, both these mechanisms could produce somewhat lower values
of
$M_U$. However, $M_U\gsim10^{16}\GeV$ is necessary to avoid too rapid proton
decay due to dimension-six operators \EKN. In these more general cases
the $SU(5)$ and $U(1)$ gauge couplings would not unify at $M_U$ (only
$\alpha_2$ and $\alpha_3$ would), although they would eventually
``superunify" at the string scale $M_{SU}\sim10^{18}\GeV$. To simplify matters,
below we consider the simplest possible case of $M_U=M_{SU}\sim10^{18}\GeV$.

We also draw inspiration from string model-building and regard the Higgs
mixing term $\mu h\bar h$ as a result of an effective higher-order coupling
\decisive, instead of as a result of a light singlet field getting a small
vev (\ie, $\lambda h\bar h\phi\to\lambda\vev{\phi}h\bar h$) as originally
considered \refs{\revitalized,\faspects}.

For the supersymmetry breaking parameters we consider the no-scale ansatz \LN,
which ensures the vanishing of the (tree-level) cosmological constant even
after supersymmetry breaking. This framework also arises in the low-energy
limit of superstring theory \Witten. In a theory which contains heavy fields,
the minimal no-scale structure $SU(1,1)$ \nsI\ is generalized to
$SU(N,1)$ \nsII\ which implies that the scalar fields do not feel the
supersymmetry breaking effects. In practice this means that the universal
scalar mass ($m_0$) and the universal cubic scalar coupling ($A$) are set to
zero. The sole source of supersymmetry breaking is the universal gaugino mass
($m_{1/2}$). We first let the universal bilinear scalar coupling ($B$) float,
\ie, be determined by the radiative electroweak symmetry breaking constraints.
We also consider the strict no-scale scenario where $B(M_U)=0$. It is worth
pointing out that within the no-scale framework the value of $m_{1/2}$ should
be determined dynamically and explicit calculations \sample\ show that
it should be below $1\TeV$. A recent analysis has shown that this result may
also occur automatically once all phenomenological constraints on the model
have been imposed \noscale.

This paper is organized as follows. In Sec. 2 we present the string-inspired
model with all the model-building details which determine in principle the
masses of the new heavy vector-like particles. We also discuss the question
of the possible re-introduction of dangerous dimension-five proton
decay operators in this generalized model. We then impose the constraint of
flipped $SU(5)$ unification and string unification to occur at
$M_U=10^{18}\GeV$
to deduce the unknown masses. In Sec. 3 we consider the experimental
predictions
for all the sparticle and one-loop corrected Higgs boson masses in this model,
and deduce several simple relations among the various sparticle masses. In
Sec. 4 we repeat this analysis for the strict no-scale case. This additional
constraint allows us to determine $\tan\beta$ for a given $m_{\tilde g}$ and
$m_t$ (up to a possible two-fold ambiguity), and thus to sharpen the most
$\tan\beta$-sensitive predictions. In Sec. 5 we summarize our conclusions.
\newsecnpb{The Model}
The model we consider is a generalization of that presented in
Ref. \revitalized, and contains the following flipped $SU(5)$ fields:
\item{(i)} three generations of quark and lepton fields $F_i,\bar f_i,l^c_i,\,
i=1,2,3$;
\item{(ii)}two pairs of Higgs \r{10},\rb{10} representations $H_i,\bar H_i,\,
i=1,2$;
\item{(iii)}one pair of ``electroweak" Higgs \r{5},\rb{5} representations
$h,\bar h$;
\item{(iv)}three singlet fields $\phi_{1,2,3}$.
\smallskip
\noindent Under $SU(3)\times SU(2)$ the various flipped $SU(5)$ fields
decompose as follows:
\eqna\I
$$\eqalignno{F_i&=\{Q_i,d^c_i,\nu^c_i\},\quad \bar f_i=\{L_i,u^c_i\},
\quad l^c_i=e^c_i,&\I a\cr
H_i&=\{Q_{H_i},d^c_{H_i},\nu^c_{H_i}\},\quad
\bar H_i=\{Q_{\bar H_i},d^c_{\bar H_i},\nu^c_{\bar H_i}\},&\I b\cr
h&=\{H,D\},\quad \bar h=\{\bar H,\bar D\}.&\I c\cr}$$
The most general effective\foot{To be understood in the string context as
arising from cubic and higher order terms \refs{\KLN,\decisive}.}
superpotential consistent with $SU(5)\times U(1)$ symmetry is given by
\eqna\II
$$\eqalignno{W&=\lambda^{ij}_1 F_iF_jh+\lambda^{ij}_2 F_i\bar f_j \bar h
+\lambda^{ij}_3 \bar f_il^c_j h +\mu h\bar h
+\lambda^{ij}_4 H_iH_jh+\lambda^{ij}_5\bar H_i\bar H_j\bar h\cr
&+\lambda^{ij}_{1'}H_iF_jh+\lambda^{ij}_{2'}H_i\bar f_j\bar h
+\lambda^{ijk}_6 F_i\bar H_j\phi_k+w^{ij}H_i\bar H_j+\mu^{ij}\phi_i\phi_j.
								&\II{}}$$
Symmetry breaking is effected by non-zero vevs $\vev{\nu^c_{H_i}}=V_i$,
$\vev{\nu^c_{\bar H_i}}=\bar V_i$, such that
$V^2_1+V^2_2=\bar V^2_1+\bar V^2_2$.
\subsecnpb{HIGGS DOUBLET AND TRIPLET MASS MATRICES}
The Higgs doublet mass matrix receives contributions from
$\mu h\bar h\to \mu H\bar H$ and $\lambda^{ij}_{2'}H_i\bar f_j\bar h\to
\lambda^{ij}_{2'}V_i L_j\bar H$. The resulting matrix is
\eqn\IIa{{\cal M}_2=\bordermatrix{&\bar H\cr H&\mu\cr
L_1&\lambda^{i1}_{2'}V_i\cr
L_2&\lambda^{i2}_{2'}V_i\cr
L_3&\lambda^{i3}_{2'}V_i\cr}.}
To avoid fine-tunings of the $\lambda^{ij}_{2'}$ couplings we must demand
$\lambda^{ij}_{2'}\equiv0$, so that $\bar H$ remains light.

The Higgs triplet matrix receives several contributions:
$\mu h\bar h\to\mu D\bar D$;
$\lambda^{ij}_{1'}H_iF_jh\to\lambda^{ij}_{1'}V_id^c_jD$;
$\lambda^{ij}_4 H_iH_jh\to\lambda^{ij}_4V_i d^c_{H_j}D$;
$\lambda^{ij}_5 \bar H_i\bar H_j\bar h
\to\lambda^{ij}_5\bar V_i d^c_{\bar H_j}\bar D$;
$w^{ij}d^c_{H_i}d^c_{\bar H_j}$. The resulting matrix is\foot{The zero entries
in ${\cal M}_3$ result from the assumption $\vev{\phi_k}=0$ in
$\lambda_6^{ijk}F_i\bar H_j\phi_k$.}
\eqn\IIb{{\cal M}_3=\bordermatrix{
&\bar D&d^c_{H_1}&d^c_{H_2}&d^c_1&d^c_2&d^c_3\cr
D&\mu&\lambda^{i1}_4V_i&\lambda^{i2}_4V_i&\lambda^{i1}_{1'}V_i
&\lambda^{i2}_{1'}V_i&\lambda^{i3}_{1'}V_i\cr
d^c_{\bar H_1}&\lambda^{i1}_5\bar V_i&w_{11}&w_{12}&0&0&0\cr
d^c_{\bar H_2}&\lambda^{i2}_5\bar V_i&w_{21}&w_{22}&0&0&0\cr}.}
Clearly three linear combinations of $\{\bar D,d^c_{H_{1,2}},d^c_{1,2,3}\}$
will remain light. In fact, such a general situation will induce a mixing
in the down-type Yukawa matrix $\lambda^{ij}_1 F_iF_jh\to\lambda^{ij}_1Q_i
d^c_jH$, since the $d^c_j$ will need to be re-expressed in terms of these
mixed light eigenstates.\foot{Note that this mixing is on top of any structure
that $\lambda^{ij}_1$ may have, and is the only source of mixing in the typical
string model-building case of a diagonal $\lambda_2$ matrix.} This low-energy
quark-mixing mechanism is an explicit realization of the general
extra-vector-abeyance (EVA) mechanism of Ref. \EVA. As a first approximation
though, in what follows we will set $\lambda^{ij}_{1'}=0$, so that the light
eigenstates are $d^c_{1,2,3}$.
\subsecnpb{NEUTRINO SEE-SAW MATRIX}
The see-saw neutrino matrix receives contributions from:
$\lambda^{ij}_2F_i\bar f_j\bar h\to m^{ij}_u\nu^c_i\nu_j$;
$\lambda^{ijk}_6 F_i\bar H_j\phi_k\to\lambda^{ijk}_6\bar V_j\nu^c_i\phi_k$;
$\mu^{ij}\phi_i\phi_j$. The resulting matrix is\foot{We neglect a possible
higher-order contribution which could produce a non-vanishing $\nu^c_i\nu^c_j$
entry \chorus.}
\eqn\IIc{{\cal M}_\nu=\bordermatrix{&\nu_j&\nu^c_j&\phi_j\cr
\nu_i&0&m^{ji}_u&0\cr
\nu^c_i&m^{ij}_u&0&\lambda^{ikj}_6\bar V_k\cr
\phi_i&0&\lambda^{jki}_6\bar V_k&\mu^{ij}\cr}.}
\subsecnpb{NUMERICAL SCENARIO}
To simplify the discussion we will assume, besides\foot{In Ref.
\revitalized\ the discrete symmetry $H_1\to -H_1$ was imposed so that these
couplings automatically vanish when $H_2,\bar H_2$ are not present. This
symmetry (generalized to $H_i\to -H_i$) is not needed here since it would
imply $w^{ij}\equiv0$, which is shown below to be disastrous for gauge
coupling unification.} $\lambda^{ij}_{1'}=\lambda^{ij}_{2'}\equiv0$, that
\eqna\IId
$$\eqalignno{\lambda^{ij}_4&=\delta^{ij}\lambda^{(i)}_4,\quad
\lambda^{ij}_5=\delta^{ij}\lambda^{(i)}_5,\quad
\lambda^{ijk}_6=\delta^{ij}\delta^{ik}\lambda^{(i)}_6,\cr
\mu^{ij}&=\delta^{ij}\mu_i,\quad w^{ij}=\delta^{ij}w_i.&\IId{}}$$
These choices are likely to be realized in string versions of this model
and will not alter our conclusions below.
In this case the Higgs triplet mass matrix reduces to
\eqn\III{{\cal M}_3=\bordermatrix
{&\bar D&d^c_{H_1}&d^c_{H_2}\cr
D&\mu&\lambda^{(1)}_4V_1&\lambda^{(2)}_4V_2\cr
d^c_{\bar H_1}&\lambda^{(1)}_5\bar V_1&w_1&0\cr
d^c_{\bar H_2}&\lambda^{(2)}_5\bar V_2&0&w_2\cr}.}
Regarding the $(3,2)$ states, the scalars get either eaten by the $X,Y$ $SU(5)$
heavy gauge bosons or become heavy Higgs bosons, whereas the fermions interact
with the $\wt X,\wt Y$ gauginos through the following mass matrix \search
\eqn\IV{{\cal M}_{(3,2)}=\bordermatrix
{&Q_{\bar H_1}&Q_{\bar H_2}&\wt Y\cr
Q_{H_1}&w_1&0&g_5V_1\cr
Q_{H_2}&0&w_2&g_5V_2\cr
\wt X&g_5\bar V_1&g_5\bar V_2&0\cr}.}
The lightest eigenvalues of these two matrices (denoted generally by $d^c_H$
and $Q_H$ respectively) constitute the new relatively light particles in the
spectrum, which are hereafter referred to as the `{\it gap}' particles since
with suitable masses they bridge the gap between unification masses at
$10^{16}\GeV$ and $10^{18}\GeV$.

Guided by the phenomenological requirement on the gap particle masses, \ie,
$M_{Q_H}\gg M_{d^c_H}$ \sism, we consider the following explicit numerical
scenario
\eqn\V{\lambda^{(2)}_4=\lambda^{(2)}_5=0,\quad
V_1,\bar V_1,V_2,\bar V_2\sim V\gg w_1\gg w_2\gg\mu,}
which would need to be reproduced in a viable string-derived model.
{}From Eq. \III\ we then get $M_{d^c_{H_2}}=M_{d^c_{\bar H_2}}=w_2$, and all
other mass eigenstates $\sim V$. Furthermore, ${\cal M}_{(3,2)}$ has a
characteristic polynomial $\lambda^3-\lambda^2(w_1+w_2)-\lambda(2V^2-w_1w_2)
+(w_1+w_2)V^2=0$, which has two roots of ${\cal O}(V)$ and one root of
${\cal O}(w_1)$. The latter corresponds to $\sim(Q_{H_1}-Q_{H_2})$ and
$\sim(Q_{\bar H_1}-Q_{\bar H_2})$. In sum then, the gap particles have masses
$M_{Q_H}\sim w_1$ and $M_{d^c_H}\sim w_2$, whereas all other heavy particles
have
masses $\sim V$.

The see-saw matrix reduces to
\eqn\Va{{\cal M}_\nu=\bordermatrix{&\nu_i&\nu^c_i&\phi_i\cr
\nu_i&0&m^i_u&0\cr
\nu^c_i&m^i_u&0&\lambda^{(i)}\bar V_i\cr
\phi_i&0&\lambda^{(i)}\bar V_i&\mu^i\cr},}
for each generation. The physics of this see-saw matrix has been
discussed recently in Ref. \chorus, where it was shown to lead to an
interesting
amount of hot dark matter ($\nu_\tau$) and an MSW-effect ($\nu_e,\nu_\mu$)
compatible with all solar neutrino data.
\subsecnpb{PROTON DECAY}
The dimension-six operators mediating proton decay in this model are highly
suppressed due to the large mass of the $X,Y$ gauge bosons
($\sim M_U=10^{18}\GeV$). Higgsino mediated dimension-five operators exist
and are naturally suppressed in the minimal model of Ref. \revitalized. The
reason for this is that the Higgs triplet mixing term
$\mu h\bar h\to \mu D\bar D$ is small ($\mu\sim M_Z$), whereas the Higgs
triplet mass eigenstates obtained from Eq. \IIb\ by just keeping the $2\times2$
submatrix in the upper left-hand corner, are always very heavy ($\sim V$).
The dimension-five mediated operators are then proportional to $\mu/V^2$ and
thus the rate is suppressed by a factor or $(\mu/V)^2\ll1$ relative to the
unsuppressed case found in the standard $SU(5)$ model.

In the generalized model presented here, the Higgs triplet mixing term is
still $\mu D\bar D$. However, the exchanged mass eigenstates are not
necessarily all very heavy. In fact, above we have demanded the existence of a
relatively light ($\sim w_1$) Higgs triplet state ($d^c_H$). In this case
the operators are proportional to $\mu\alpha_i\bar\alpha_i/{\cal M}^2_i$, where
${\cal M}_i$ is the mass of the $i$-th exchanged eigenstate and
$\alpha_i,\bar\alpha_i$ are its $D,\bar D$ admixtures. In the scenario
described
above, the relatively light eigenstates ($d^c_{H_2},d^c_{\bar H_2}$) contain
no $D,\bar D$ admixtures, and the operator will again be $\propto\mu/V^2$.

Note however that if conditions \V\ (or some analogous suitability requirement)
are not satisfied, then diagonalization of ${\cal M}_3$ in Eq. \III\ may
re-introduce a sizeable dimension-five mediated proton decay rate, depending
on the value of the $\alpha_i,\bar\alpha_i$ coefficients. To be safe one
should demand \refs{\dfive,\LNZ}
\eqn\Vb{{\mu\alpha_i\bar\alpha_i\over {\cal M}^2_i}\lsim{1\over 10^{17}\GeV}.}
For the higher values of $M_{d^c_H}$ in Table I (see below), this constraint
can be satisfied for not necessarily small values of $\alpha_i,\bar\alpha_i$.
\subsecnpb{GAUGE COUPLING UNIFICATION}
Since we have chosen $V\sim M_U=M_{SU}=10^{18}\GeV$, this means that the
Standard Model gauge couplings should unify at the scale $M_U$. However, their
running will be modified due to the presence of the gap particles. Note that
the underlying flipped $SU(5)$ symmetry, even though not evident in this
respect, is nevertheless essential in the above discussion. The masses $M_Q$
and $M_{d^c_H}$ can then be determined, as follows \sism
\eqna\VI
$$\eqalignno{\ln{M_{Q_H}\over
m_Z}=&\pi\left({1\over2\alpha_e}-{1\over3\alpha_3}
-{\sin^2\theta_w-0.0029\over\alpha_e}\right)-2\ln{M_U\over m_Z}-0.63,&\VI a\cr
\ln{M_{d^c_H}\over m_Z}=&\pi\left({1\over2\alpha_e}-{7\over3\alpha_3}
+{\sin^2\theta_w-0.0029\over\alpha_e}\right)-6\ln{M_U\over m_Z}-1.47,&\VI
b\cr}$$
where $\alpha_e$, $\alpha_3$ and $\sin^2\theta_w$ are all measured at $M_Z$.
This is a one-loop determination (the constants account for the dominant
two-loop corrections) which neglects all low- and high-energy threshold
effects,\foot{Here we assume a common supersymmetric threshold at $M_Z$.
In fact, the supersymmetric threshold and the $d^c_H$ mass are anticorrelated.
See Ref. \sism\ for a discussion.} but is quite adequate for our present
purposes. As shown in Table I (and formula \VI{b}) the $d^c_H$ mass depends
most sensitively on
$\alpha_3(M_Z)=0.118\pm0.008$ \Bethke, whereas the $Q_H$ mass and the unified
coupling are rather insensitive to it. The unification of the gauge couplings
is shown in Fig. 1 (solid lines) for the central value of $\alpha_3(M_Z)$. This
figure also shows the case of no gap particles (dotted lines), for which
$M_U\approx10^{16}\GeV$.

\topinsert
\hrule\smallskip
\noindent{\bf Table I}: The value of the gap particle masses and the unified
coupling for $\alpha_3(M_Z)=0.118\pm0.008$. We have taken $M_U=10^{18}\GeV$,
$\sin^2\theta_w=0.233$, and $\alpha^{-1}_e=127.9$.
\smallskip
\input tables
\thicksize=1.0pt
\centerjust
\begintable
$\alpha_3(M_Z)$|$M_{d^c_H}\,(\GeV)$|$M_{Q_H}\,(\GeV)$|$\alpha(M_U)$\cr
$0.110$|$4.9\times10^4\GeV$|$2.2\times10^{12}\GeV$|$0.0565$\nr
$0.118$|$4.5\times10^6\GeV$|$4.1\times10^{12}\GeV$|$0.0555$\nr
$0.126$|$2.3\times10^8\GeV$|$7.3\times10^{12}\GeV$|$0.0547$\endtable
\smallskip
\hrule\medskip
\endinsert
\newsecnpb{Experimental Predictions}
The model presented in the previous section can be analyzed to determine its
low-energy experimental predictions for \eg, the Higgs and sparticle masses.
Consistent with the assumption of flipped $SU(5)$ gauge symmetry breaking
at $\sim M_U=10^{18}\GeV$, we assume that the onset of universal supersymmetry
breaking in the observable sector occurs at this same scale \faspects. This
can be parametrized in terms of a universal gaugino mass ($m_{1/2}$), a
universal scalar mass ($m_0$), and universal trilinear ($A$) and bilinear
($B$) scalar couplings. One also needs to specify the fermion Yukawa couplings
and the Higgs mixing parameter $\mu$. The renormalization group equations then
run the relevant parameters to low energies where radiative electroweak
symmetry breaking occurs (studied using the one-loop effective potential).
When all is said and done, the whole theory can be specified in terms of just
five parameters: $m_{1/2},m_0,A$, the ratio of Higgs vacuum expectation values
$\tan\beta$, and the top-quark mass $m_t$. Note that in this scheme $\mu$ and
$B$ are calculated quantities; the sign of $\mu$ remains undetermined. Our
calculations enforce all known experimental bounds on supersymmetric and
one-loop corrected Higgs masses. We refer the reader to Ref. \aspects\ for
a detailed account of this procedure. As discussed in the Introduction, in what
follows we consider the typical no-scale supergravity boundary conditions \LN,
where $m_0=A=0$.\foot{In Refs. \refs{\noscale,\aspects} a similar analysis was
performed for a model without the gap particles (\ie, where
$M_U\sim10^{16}\GeV$). In Ref. \aspects\ an $SU(3)\times
SU(2)\times U(1)$ version of the model presented in this paper was considered
(referred to as the SISM model), although only a rather
limited analysis was performed.} In this section we let $B$ float and in the
following section we consider the strict no-scale case where $B(M_U)=0$ is
required.

For each sign of $\mu$ we have explored a three-dimensional grid in this
parameter space: $\tan\beta=2-50\,(2)$, $m_{1/2}=50-500\,(6)\GeV$,
$m_t=95-195\,(5)\GeV$, where the numbers in parenthesis indicate the size of
the step taken in that particular direction. Larger values of $\tan\beta$
and/or $m_t$ violate perturbative unification, and $m_{1/2}>500\GeV$ leads to
$m_{\tilde q},m_{\tilde g}>1\TeV$, which would make the theory
``unnatural". As discussed in the Introduction, the correct superstring model
will have to provide an explanation for why these masses are light, and if not
so, why this is not unnatural. For now we just take this to be true realizing
that relaxing this assumption will not add regions of parameter space which
could be tested experimentally at the next generation of colliders. On the
other hand, at least in the realm of supergravity, the condition $m_{\tilde q},
m_{\tilde g}<1\TeV$ is granted by the no-scale supergravity mechanism \sample\
and we believe that the correct superstring model will reproduce this important
condition. Our exploration resulted in $\approx12$K acceptable
points for each sign of $\mu$, and for all of these we found
\eqn\VII{\tan\beta\lsim32\quad{\rm and}\quad m_t\lsim185\GeV.}
\subsecnpb{MASS RANGES}
The restriction of $m_{\tilde q},m_{\tilde g}<1\TeV$ cuts off the growth of
most of the sparticle and Higgs masses at $\approx1\TeV$. However, the
sleptons, the lightest Higgs, the two lightest neutralinos, and the lightest
chargino are cut off at a much lower mass, as follows\foot{In this class of
supergravity models the three sneutrinos ($\tilde\nu$) are degenerate in mass.
Also, $m_{\tilde \mu_L}=m_{\tilde e_L}$ and $m_{\tilde\mu_R}=m_{\tilde e_R}$.}
\eqna\VIII
$$\eqalignno{m_{\tilde e_R}&<190\GeV,\quad m_{\tilde e_L}<305\GeV,
\quad m_{\tilde\nu}<295\GeV,\quad m_{\tilde\tau_1}<185\GeV,
\quad m_{\tilde\tau_2}<315\GeV,\cr
m_h&<135\GeV,\quad m_{\chi^0_1}<145\GeV,\quad m_{\chi^0_2}<285\GeV,
\quad m_{\chi^\pm_1}<285\GeV.&\VIII{}\cr}$$
It is interesting to note that due to the various constraints on the model,
the gluino and squark masses are predicted to satisfy the current experimental
bounds automatically. We find $m_{\tilde g}\gsim220\GeV$ and
$m_{\tilde q}\gsim200\GeV$, except for the lightest stop eigenstate
$\tilde t_1$, which can be as light as $\approx150\GeV$. Therefore, the
$\tilde t_1$ squark could be the first squark to be possibly observed at
Fermilab in the near future.
\subsecnpb{MASS RELATIONS}
The first and second generation squark and slepton masses can be determined
analytically
\eqn\IX{\wt m^2_i=m^2_{1/2}(c_i+\xi^2_0)-d_i{\tan^2\beta-1\over\tan^2\beta+1}
M^2_W,}
where $d_i=(T_{3i}-Q)\tan^2\theta_w+T_{3i}$ (\eg, $d_{\tilde u_L}={1\over2}
+{1\over6}\tan^2\theta_w$, $d_{\tilde e_R}=-\tan^2\theta_w$), and in our case
$\xi_0=m_0/m_{1/2}=0$. The coefficients
$c_i$ can be calculated numerically in terms of the low-energy gauge couplings,
and are given in Table II\foot{These are renormalized at the scale $M_Z$. In
a more accurate treatment, the $c_i$ would be renormalized at the physical
sparticle mass scale, leading to second order shifts on the sparticle masses.}
for $\alpha_3(M_Z)=0.118\pm0.008$. In the table it
is also shown $c_{\tilde g}=m_{\tilde g}/m_{1/2}$. The ``average" squark
mass, $m_{\tilde q}\equiv{1\over8}(m_{\tilde u_L}+m_{\tilde u_R}
+m_{\tilde d_L}+m_{\tilde d_R}+m_{\tilde c_L}+m_{\tilde c_R}
+m_{\tilde s_L}+m_{\tilde s_R})$ is then determined to be
\eqn\X{m_{\tilde q}=0.97 m_{\tilde g},}
within $\pm3\%$, allowing for a $\pm1\sigma$ error in $\alpha_3(M_Z)$ (the
dependence
on $\tan\beta$ is negligible). The squark splitting around the average is
$\approx2\%$.

\topinsert
\hrule\smallskip
\noindent{\bf Table II}: The value of the $c_i$ coefficients appearing in
Eq. \IX\ for $\alpha_3(M_Z)=0.118\pm0.008$. Also shown is the ratio
$c_{\tilde g}=m_{\tilde g}/m_{1/2}$.
\smallskip
\input tables
\thicksize=1.0pt
\centerjust
\begintable
$i$|$c_i\,(0.110)$|$c_i\,(0.118)$|$c_i\,(0.126)$\cr
$\tilde u_L,\tilde d_L$|$3.98$|$4.41$|$4.97$\nr
$\tilde u_R$|$3.68$|$4.11$|$4.66$\nr
$\tilde d_R$|$3.63$|$4.06$|$4.61$\nr
$\tilde\nu,\tilde e_L$|$0.406$|$0.409$|$0.413$\nr
$\tilde e_R$|$0.153$|$0.153$|$0.153$\nr
$c_{\tilde g}$|$1.95$|$2.12$|$2.30$\endtable
\smallskip
\hrule\medskip
\endinsert

The third-generation squarks deviate considerably from the average squark mass
and have a non-negligible dependence on $\tan\beta$ due to the off-diagonal
elements on the squark mass matrix (which are proportional to the corresponding
quark mass). Throughout the parameter space we found the following maximal
relative deviations of these squark masses relative to the average squark
mass (\ie, $|m_{\tilde q_i}-m_{\tilde q}|/m_{\tilde q}$):
\eqn\XI{\tilde b_1:\lsim14\%;\quad \tilde b_2:\lsim8\%;
\quad \tilde t_1:\lsim47\%;\quad\tilde t_2:\lsim35\%.}
In Fig. 2 we plot\foot{For all the scatter plots shown in this paper we have
restricted the values of the top-quark mass to $m_t=100,130,160\GeV$ to have a
manageable number of points.} the sbottom and stop masses. The sbottom masses
are not split enough so that $m_{\tilde b_1}$ and $m_{\tilde b_2}$ blend into
a wide band. The stop masses are separated in the figure. Note the lesser
definition of the $\tilde t_1$ masses due to the relatively larger $m_t$ and
$\tan\beta$ effects. For all these masses one should note that the average
squark mass $m_{\tilde q}=0.97m_{\tilde g}$ runs somewhere in between
these mass bands.

The sleptons are much lighter than the squarks since roughly
$m_{\tilde l}/m_{\tilde q}\approx(c_{\tilde l}/c_{\tilde q})^{1/2}\lsim0.3$.
In principle, for small $m_{\tilde g}$, one would expect a stronger $\tan\beta$
dependence for sleptons due to the relatively smaller contribution of the first
term in Eq. \IX. This, together with the large difference between
$c_{\tilde\nu,\tilde e_L}$ and $c_{\tilde e_R}$, implies that an ``average"
slepton mass (as usually assumed in phenomenological studies of the MSSM) is a
rather {\it poor} approximation to this model. In Fig. 3 we show the
$\tilde\tau_{1,2}$ and $\tilde e_{L,R}$ masses; the inadequacy of the average
slepton mass approximation is evident. As expected, the off-diagonal elements
in the $\tilde\tau$ mass matrix give a broad band of $\tilde\tau_{1,2}$ masses
for a given $m_{\tilde g}$ value. On the other hand, the $\tilde e_{L,R}$
masses look much sharper as a function of $m_{\tilde g}$. What happens is
that for small $m_{\tilde g}$, when the $\tan\beta$ effects are potentially
important, $\tan\beta$ is not allowed to become large and thus the $D$-term
is suppressed. The $\tilde\nu$ masses start off below $m_{\tilde e_R}$ and
quickly approach the $m_{\tilde e_L}$ line. In numbers we find
\eqn\XII{m_{\tilde e_L}\approx m_{\tilde\nu}\approx0.302\,m_{\tilde g};
\quad m_{\tilde e_R}\approx0.185\,m_{\tilde g},}
where the small $D$-term contribution has been neglected; it becomes
negligible for increasingly larger values of $m_{\tilde g}$. The $\tilde\tau_1$
($\tilde\tau_2$) mass approximates $\tilde e_R$ ($\tilde e_L,\tilde\nu$) as
a ``central value", but has quite a spread around it, as Fig. 3 shows.

We find that throughout the parameter space $|\mu|$ is generally much larger
than $M_W$ and $|\mu|>M_2$. This is shown on the top row of Fig. 4. Note
that $|\mu|\propto m_{\tilde g}$ with the $\tan\beta$-dependent slope growing
with the value of $m_t$ \aspects; the three values of $m_t$ used are evident
in the figure. This behavior points to a simple eigenvalue structure
for the two lightest neutralinos and the lightest chargino \ANc, as follows
\eqn\XIII{m_{\chi^0_1}\approx\coeff{1}{2}m_{\chi^0_2};\quad
m_{\chi^0_2}\approx m_{\chi^\pm_1}\approx M_2=\alpha_2/\alpha_3 m_{\tilde g}
\approx0.3m_{\tilde g}.}
In practice we find $m_{\chi^0_2}\approx m_{\chi^\pm_1}$ to be satisfied
quite accurately (see Fig. 5, top row), whereas $m_{\chi^0_1}\approx{1\over2}
m_{\chi^0_2}$ is only qualitatively satisfied (see Fig. 5, bottom row).
In fact, these two mass relations are much more reliable than the one that
links them to $m_{\tilde g}$ (not shown). The heavier neutralino
($\chi^0_{3,4}$) and chargino ($\chi^\pm_2$) masses are determined by the
value of $|\mu|$ (shown in Fig. 4); they all approach this limit for large
enough $|\mu|$. More precisely, $m_{\chi^0_3}$ approaches $|\mu|$ sooner than
$m_{\chi^0_4}$ does. On the other hand, $m_{\chi^0_4}$ approaches
$m_{\chi^\pm_2}$ rather quickly.

The one-loop corrected lightest Higgs boson mass ($m_h$) is shown in Fig. 6
(top row). The three noticeable bands correspond to the three values of the
top-quark mass used ($m_t=100,130,160\GeV$), for large values of $\tan\beta$.
For smaller values of $\tan\beta$ the tree-level contribution is suppressed
and the curves reach down to low values of $m_h$. In this case one can most
easily note the expected logarithmic rise of $m_h$ with the squark mass
(recall that $m_{\tilde q}\approx m_{\tilde g}$). For low $m_t$ the curves
rise very slightly. However, for large $m_t$ these rise quite dramatically.
This is all in agreement with the expected behavior deduced from the
approximate analytical expressions for $m_h$ in the literature. In the figure
($m_t\le160\GeV$) we get $m_h\lsim120\GeV$. Allowing for the whole range of
$m_t$ values this upper bound gets relaxed to $\approx135\GeV$. The one-loop
corrected pseudoscalar mass $m_A$ is also shown in Fig. 6 as a function of
$m_h$. The three bands correspond to from-left-to-right $m_t=100,130,160\GeV$.
The predictions for $m_A$ are not very sharp. It is nevertheless true that
for all points examined $m_A>m_h$, as expected from general considerations
\DNh. The other two Higgs boson states, $H$ and $H^+$, are approached from
below by $m_A$ \higgs. For $m_{H,H^+}\gsim200\,(300)\GeV$ the difference is
$\lsim8\%\,(3\%)$.

To appreciate the relations among the sparticle masses in this model, in
Fig. 7 we show a graphical display of the spectrum for $m_{\tilde g}=300\GeV$
and $m_t=130\GeV$ and both signs of $\mu$. The masses generally scale with
$m_{\tilde g}$. The masses shown are also given in
Table III where in addition we give the percentage deviations of the masses
relative to their central values due to the variation of $\tan\beta$ over
all its allowed range ($2\lsim\tan\beta\lsim32$).
\topinsert
\hrule\smallskip
\noindent{\bf Table III}: Central values of the sparticle and one-loop
corrected Higgs boson masses for $\alpha_3(M_Z)=0.118$, $m_{\tilde g}=300\GeV$,
$m_t=130\GeV$, and both signs of $\mu$, showing the percentage deviations from
the central value due to the variation of $\tan\beta$ over its whole allowed
range ($2\lsim\tan\beta\lsim32$).
\smallskip
\input tables
\thicksize=1.0pt
\centerjust
\begintable
$i$\|$\widetilde m_i(\mu>0)$|\%\|$\widetilde m_i(\mu<0)$|\%\cr
$\mu$\|$164$|$17$\|$-164$|$17$\nr
$\tilde q$\|$290$|$0.1$\|$290$|$0.1$\nr
$\chi^0_1$\|$43$|$8.9$\|$29$|$18$\nr
$\chi^0_2$\|$83$|$19$\|$62$|$2.1$\nr
$\chi^0_3$\|$178$|$12$\|$174$|$12$\nr
$\chi^0_4$\|$197$|$11$\|$212$|$13$\nr
$\chi^\pm_1$\|$83$|$20$\|$53$|$9.3$\nr
$\chi^\pm_2$\|$201$|$8.2$\|$210$|$11$\nr
$h$\|$79$|$23$\|$82$|$19$\nr
$H$\|$183$|$35$\|$185$|$32$\nr
$A$\|$177$|$32$\|$179$|$30$\nr
$H^+$\|$195$|$26$\|$198$|$24$\nr
$\tilde e_L$\|$100$|$2.2$\|$100$|$2.2$\nr
$\tilde e_R$\|$68$|$4.2$\|$68$|$4.2$\nr
$\tilde\nu$\|$69$|$8.5$\|$69$|$8.5$\nr
$\tilde\tau_1$\|$56$|$21$\|$57$|$18$\nr
$\tilde\tau_2$\|$105$|$7.3$\|$105$|$6.8$\nr
$\tilde b_1$\|$270$|$3.5$\|$267$|$3.2$\nr
$\tilde b_2$\|$286$|$0.5$\|$290$|$1.0$\nr
$\tilde t_1$\|$205$|$8.2$\|$170$|$7.8$\nr
$\tilde t_2$\|$344$|$3.4$\|$363$|$1.3$\endtable
\smallskip
\hrule\medskip
\endinsert
\subsecnpb{NEUTRALINO DARK MATTER}
In Fig. 4 (bottom row) we plot the result for the cosmic relic abundance of the
lightest neutralino, $\Omega_\chi h^2_0$. This has been calculated following
the
methods of Refs. \LNY. Since $m_\chi\equiv m_{\chi^0_1}$ grows with
$m_{\tilde g}$, and $|\mu|$ grows with $m_{\tilde g}$, then as $m_\chi$ grows,
the pure gaugino region ($|\mu|\gg M_2\approx0.3m_{\tilde g}$) is approached
and the neutralino pair-annihilation is suppressed, leading to larger
$\Omega_\chi h^2_0$ values. Note the effect of the $Z$-pole for
$m_\chi\approx{1\over2}M_Z$. We find
that $\Omega_\chi h^2_0$ can be as large as $\approx0.25$. This result is in
good agreement with the observational upper bound on $\Omega_\chi h^2_0$ \KT\
and does not constrain the model any further. Moreover, fits to the COBE data
and the
small and large scale structure of the Universe suggest \many\ a mixture
of $\approx70\%$ cold dark matter and $\approx30\%$ hot dark matter together
with $h_0\approx0.5$. The hot dark matter component in the form of massive
tau neutrinos has already been shown to be compatible with the flipped $SU(5)$
model we consider here \chorus, whereas the cold dark matter component implies
$\Omega_\chi h^2_0\approx0.17$ which is reachable in this model for
$m_\chi\gsim100\GeV$.

It is interesting to note that values of $\Omega_\chi h^2_0\lsim0.25$ occur
naturally in this model, and in general for $m_{1/2}\gg m_0$. This situation is
in sharp contrast with for example, the minimal $SU(5)$ supergravity model,
where $\Omega_\chi h^2_0\gg1$ occurs naturally instead \LNZ.
\newsecnpb{The strict no-scale case}
We now impose the additional constraint on the theory that $B(M_U)=0$, that
is the strict no-scale case. Since $B(M_Z)$ is determined by the radiative
electroweak symmetry breaking conditions, this added constraint needs to be
imposed in a rather indirect way. That is, for given $m_{\tilde g}$ and $m_t$
values, we scan the possible values of $\tan\beta$ looking for cases where
$B(M_U)=0$. The most striking result is that solutions exist {\it only} for
$m_t\lsim135\GeV$ if $\mu>0$ and for $m_t\gsim140\GeV$ if $\mu<0$. That is,
the value of $m_t$ {\it determines} the sign of $\mu$. Furthermore, for $\mu<0$
the value of $\tan\beta$ is
determined uniquely as a function of $m_t$ and $m_{\tilde g}$, whereas for
$\mu>0$, $\tan\beta$ can be double-valued for some $m_t$ range which includes
$m_t=130\GeV$ but does not include $m_t=100\GeV$. In Fig. 8 (top row) we plot
the solutions found in this manner for the indicated $m_t$ values.

All the mass relationships deduced in the previous section apply here as well.
The $\tan\beta$-spread that some of them have will be much reduced though.
The most noticeable changes occur for the quantities which depend most
sensitively on $\tan\beta$, \ie, the neutralino relic abundance and the
lightest and pseudoscalar Higgs masses. In Fig. 8 (bottom row) we plot
$\Omega_\chi h^2_0$ versus $m_\chi$ for this case. Note that continuous
values of $m_t$ will tend to fill in the space between the lines shown.
In Fig. 9 (top row) we plot the one-loop corrected lightest Higgs boson
mass versus $m_{\tilde g}$. The result is that $m_h$ is basically determined
by $m_t$; only a weak dependence on $m_{\tilde g}$ exists. Moreover, for
$m_t\lsim135\GeV\Leftrightarrow\mu>0$, $m_h\lsim105\GeV$; whereas for
$m_t\gsim140\GeV\Leftrightarrow\mu<0$, $m_h\gsim100\GeV$. Therefore, in the
strict no-scale case, once the top-quark mass is measured, we will know the
sign of $\mu$ and whether $m_h$ is above or below $100\GeV$. The pseudoscalar
Higgs mass dependence on $m_h$ is also much simplified, as Fig. 9 (bottom row)
shows.

For $\mu>0$, we just showed that the strict no-scale constraint requires
$m_t\lsim135\GeV$. This implies that $\mu$ cannot grow as large as it did
previously. In fact, for $\mu>0$, $\mu_{max}\approx745\GeV$ before and
$\mu_{max}\approx440\GeV$ now. This smaller value of $\mu_{max}$ has the
effect of cutting off the growth of the $\chi^0_{3,4},\chi^\pm_2$ masses
at $\approx\mu_{max}\approx440\GeV$ (c.f. $\approx750\GeV$) and of the heavy
Higgs masses at $\approx530\GeV$ (c.f. $\approx940\GeV$).

\newsecnpb{Conclusions}
In this paper we have presented the simplest, string-derivable, supergravity
model and deduced its experimental predictions. This new string-inspired model
has several
features that are found in real string-derived models, such as string
unification and a unified gauge group which can reduce to the Standard Model
one after spontaneous gauge symmetry breaking. We also demanded that the
low-energy supergravity theory be of the no-scale type, since this general
framework is supported by superstring theory. The model built this way
should be considered to be an idealization of what its string-derived
incarnation should be. In the process we have identified several potential
model-building problems which would need to be watched for in a string
implementation. We have assumed that the various needed mass scales are
generated somehow and have fit their values to achieve string unification
at $M_U=10^{18}\GeV$. The actual origin of these mass scales will lie within
the structure of the successful string model. Known examples include
condensates \refs{\decisive,\conden} and vacuum expectation values \ART\
of hidden matter fields. As in any non-minimal flipped $SU(5)$ model,
non-negligible dimension-five proton decay operators could be re-introduced.
In the model presented here these remain highly suppressed. However, in
variants of this model or in string-derived versions, these operators could
exist at an observable level. This question deserves further study.

We have also performed a thorough and accurate exploration of the parameter
space of the model and solved for all the sparticle and one-loop corrected
Higgs masses. The growth of the supersymmetry breaking parameter is cut-off
by the no-scale supergravity mechanism which guarantees
$m_{\tilde q},m_{\tilde g}<1\TeV$. We found
some general results and upper bounds on the sleptons and lightest
neutralino, chargino, and Higgs masses. We have also found several simple
relations among squark and gluino masses, among slepton masses, and among
the lightest neutralino and chargino masses. The neutralino relic abundance
$\Omega_\chi h^2_0$ never exceeds $\approx0.25$ and therefore does not
constrain this model. However, it may constitute a significant portion of
the dark matter in the Universe in general and in the galactic halo in
particular. In the strict no-scale case we find a striking result: if $\mu>0$,
$m_t\lsim135\GeV$, whereas if $\mu<0$, $m_t\gsim140\GeV$. Therefore the value
of $m_t$ determines the sign of $\mu$. Moreover, the value of $\tan\beta$ can
also be determined. Furthermore, we found that the value of $m_t$ also
determines  whether the lightest Higgs boson is above or below $100\GeV$.

\bigskip
\bigskip
\noindent{\it Acknowledgments}: This work has been supported in part by DOE
grant DE-FG05-91-ER-40633. The work of J.L. has been supported by an SSC
Fellowship. The work of D.V.N. has been supported in part by a grant from
Conoco Inc. We would like to thank the HARC Supercomputer Center for the use
of their NEC SX-3 supercomputer.
\listrefs
\listfigs
\bye